# Deciphering the Enigma of Cu-Doped Lead Apatite (LK-99): Structural Insights, Electronic Properties, and Implications for Ambient-Pressure Superconductivity

Jun Li[1] and Qi An[1,*]

[1]*Department of Materials Science and Engineering, Iowa State University, Ames, Iowa 50011, United States*

[*] Corresponding author-Email: qan@iastate.edu

## Abstract

The most recent discovery, the Cu-doped lead apatite "LK-99", is a proposed room-temperature superconductor operating under ambient pressure. However, this discovery has brought a slew of conflicting results from different scientific groups. While some observed the absence of electrical resistance, others could not confirm any signs of superconductivity in LK-99. Here, we investigate the structural and electronic properties of LK-99 and its antecedent compounds through quantum mechanics (QM) and QM-based molecular dynamics (QM-MD) simulations. Our study elucidates the insulating nature of base compounds, $Pb_{10}(PO_4)_6O$ and $Pb_{10}(PO_4)_6(OH)_2$, spotlighting their large band gaps. Notably, Cu doping in LK-99 disrupts its symmetry, yielding a distorted ground-state crystal structure with a triclinic $P1$ symmetry and $CuO_4$ square coordination. Such alterations predispose LK-99 to exhibit semiconducting behaviors, characterized by a flat band above the Fermi energy, arising from Cu-3$d$ and O-2$p$ orbitals. In addition, the S doping sustains the triclinic $P1$ symmetry but leads to a significantly reduced band gap, with a band emerging primarily from Cu-3$d$ and S-3$p$ orbitals. These findings are important in understanding LK-99's structural and electronic properties and provide a strategic compass for the development of high-T$_C$ superconductors.

**Main Text**

High-critical temperature (High-T$_C$) superconductors are often considered as the holy grail in condensed matter physics due to their great potential to revolutionize numerous technologies, including electronic communications and power energy.[1] This potential stems from their distinctive properties: zero electrical resistance and perfect diamagnetism.[1] Ever since Onnes's discovery of superconductivity in 1911,[2] scientists globally have pursued high-T$_C$ superconductors with relentless determination, leading to significant advancement in this field. Notable discoveries over the past century include the non-hydrogen BCS superconductor MgB$_2$ with a T$_C$ of 39 K,[3] and the Cu-based unconventional superconductor in the Hg–Ba–Ca–Cu–O system, which achieves a T$_C$ of 133 K[4] under ambient pressure. High-pressure examples include LaH$_{10}$ (T$_C$ = 250 K at pressure of ~170 GPa)[5] and sulfur hydride system (T$_C$ = 203 K at pressure of ~50 GPa).[6] However, the comparatively low transition temperatures of most ambient pressure superconductors, combined with the necessity for high pressure in others, remain significant barriers to the broader application of these materials. Consequently, the search for room-temperature superconductors that operate under ambient pressure continues to be an outstanding challenge in this field.

Very recently, Lee et al[7,8] reported the successful synthesis of a room-temperature superconductor, Cu-doped lead apatite, denoted as "LK-99", under ambient pressure. This announcement has elicited considerable enthusiasm within the physics community and beyond. In the wake of this claim, many independent groups have attempted to replicate the synthesis of LK-99 and its purported superconductivity, yielding a spectrum of results.[9–15] Some experiments have verified the absence of electronic resistance in LK-99 below 100 K at ambient pressure, though without observing the Meissner effect.[9] Others have detected pronounced diamagnetism, sidestepping resistivity measurement.[10] Conversely, several studies have not discerned any indications of superconductivity or diamagnetism, identifying LK-99's behavior as either semiconducting, insulating, or paramagnetic.[11–15] Notably, a potential explanation for the observed resistivity reduction in LK-99 might be the first-order structural phase transition of $Cu_2S$, transitioning from its high-temperature β phase to its low-temperature γ phase.[16,17]

To decipher the enigma of LK-99, several theoretical analyses have delved into its electronic properties and transport characteristics using quantum mechanics (QM) based density functional theory (DFT) simulations.[18–24] Contrary to the insulating behavior of lead apatite, Cu doping introduces correlated flat electronic bands of Cu-$d$ character at the Fermi level, a hallmark often seen in high-$T_C$ superconductors.[18–22] Notably, the Cu-doped configuration observed experimentally appears to be energetically non-preferred in QM assessments,[22,25–27] hinting at ambiguities in the ground-state crystalline structure of Cu-doped lead apatite.[28,29] Dynamical mean field theory (DMFT) calculations suggest that Cu-doped lead apatite manifests Mott insulating properties, driven by a pronounced interaction versus bandwidth ratio ($U/W \approx 30 \sim 50$); thus, further electron doping seems requisite for a conductive state.[23,24] Additionally, DFT explorations have contemplated doping lead apatite with various metals, including Ni, Zn, Ag, and Au, to modulate its electronic behavior.[19] Despite these efforts, critical questions remain regarding the potential superconductivity of Cu-doped lead apatite, underscoring the pressing need for deeper probes into its crystal structure, electronic structure, and other intrinsic properties.

Lead apatite typically presents in two forms: $Pb_{10}(PO_4)_6O$ and $Pb_{10}(PO_4)_6(OH)_2$,[30] which can interconvert depending on dehydration temperatures.[26] To systematically investigate the properties of Cu-doped lead apatite in the present study, we consider both of them as the parent compounds. Through QM and quantum mechanics molecular dynamics (QM-MD) simulations, this study reveals a highly distorted ground-state crystal structure for Cu-doped lead apatite. This distortion leans towards the triclinic $P1$ symmetry — a more energetically favorable state — over the trigonal $P3$ symmetry. Notably, interactions stemming from partially filled $d$-orbitals contort the $CuO_6$ octahedron into a $CuO_4$ square, characterized by four-fold coordination. This alteration in crystal symmetry paves the way for an isolated and spin-polarized empty flat band within the band gap, indicating semiconducting-like behaviors for Cu-doped lead apatite. Additionally, traces of S atoms may be embedded in the LK-99 phase, due to $Cu_2S$ impurities detected in the synthesized LK-99 samples.[7,8,17] We thus evaluate an extreme doping case, $Pb_9Cu(PO_3S)_6O$, to elucidate the influence of S atoms doping on LK-99. Intriguingly, S-doped LK-99 continues to favor the triclinic $P$1 symmetry, showcasing an isolated flat band elevated above the Fermi level and accompanied by a much narrower band gap. These findings offer profound insights into both the intrinsic properties of Cu-doped lead apatite and the broader landscape of superconductors.

To derive the ground-state crystal structure of Cu-doped lead apatite, we performed both QM and QM-MD simulations, based on DFT, via the Vienna ab initio Simulation Package (VASP).[31–33] Initially, QM-MD simulations were conducted at 300 K and ambient pressure using the isothermal-isobaric (NPT) ensemble for a 2 ps duration. During this process, the Langevin thermostat and Parinello-Rahman barostat governed the temperature and pressure, respectively. This approach ensures the lattice structure navigates beyond the potential energy surface's local minima. Subsequently, QM simulations were employed to optimize the structure in terms of lattice parameters and atomic positions. To account for the exchange-correlation interaction of electrons, the Perdew-Burke-Ernzerhof (PBE) functional form for solid (PBEsol) of the generalized gradient approximation (GGA)[34] was used. A 4 eV Hubbard-U correction was designated for Cu-$d$ states, capturing static electronic correlation effects via the Dudarev scheme.[35,36] To consider the possible spin effect, the spin polarization was considered for the calculation of both LK-99 and S-doped LK-99. Comprehensive details of the QM and QM-MD simulations are elaborated upon in the Supporting Information (SI).

In this study, we examine both $Pb_{10}(PO_4)_6O$ and $Pb_{10}(PO_4)_6(OH)_2$ as potential precursors for Cu-doped lead apatite, given that the synthesized "LK-99" sample might consist of both phases.[26]

The inherent crystal structure of these compounds, $Pb_{10}(PO_4)_6O$ and $Pb_{10}(PO_4)_6(OH)_2$, displays a hexagonal $P6_3/m$ symmetry with an associated space group 176. As shown in Figure 1a and Figure 1c, two distinct Pb sites, labeled Pb1 and Pb2, emerge due to their disparate Wyckoff sites, which are located at the interstitial space among the $PO_4$ tetrahedrons. Pb1 atoms are positioned at the $4f$ site, shaping a hexagon (illustrated by the green solid line in Figure 1), whereas Pb2 atoms sit at the $6h$ site, forming two counter-oriented triangles (highlighted by the blue solid line in Figure 1). Viewing along the $c$-axis, Pb2 atoms and adjacent $PO_4$ tetrahedrons form a cylindrical column centered around either O (O2) atoms — having a 1/4 occupancy for $Pb_{10}(PO_4)_6O$ (shown in Figure 1a) — or a chain formed by OH with 1/2 occupancy in the case of $Pb_{10}(PO_4)_6(OH)_2$ (depicted in Figure 1c). After structural optimization, the predicted lattice parameters for $Pb_{10}(PO_4)_6O$ are $a = 9.833$ Å, $c = 7.359$ Å, leading to a volume of $V = 616.14$ Å$^3$. While $Pb_{10}(PO_4)_6(OH)_2$ possesses $a = 9.886$ Å, $c = 7.383$ Å, leading to a volume of $V = 624.85$ Å$^3$. These results align closely with the experimental values of $a = 9.865$ Å, $c = 7.431$ Å, and $V = 626.29$ Å$^3$.[8] Noteworthy, the resultant symmetry for the optimized structures of $Pb_{10}(PO_4)_6O$ and $Pb_{10}(PO_4)_6(OH)_2$ transitions to P6 and $P6_3$, respectively, a consequence of the reflection disruption stemming from O atom or OH molecule alignment.[22]

Figure 1b and Figure 1d present the calculated electronic band structure and density of states (DOS) for $Pb_{10}(PO_4)_6O$ and $Pb_{10}(PO_4)_6(OH)_2$, respectively. These results indicate that both $Pb_{10}(PO_4)_6O$ and $Pb_{10}(PO_4)_6(OH)_2$ function as non-magnetic insulators, exhibiting indirect band gaps of 2.70 eV and 3.50 eV, respectively. This insulating behavior of lead apatite aligns with experimental findings.[7,8] Importantly, the partial occupancy of O2 atoms or OH molecules critically influences the band gap of lead apatite.[20] Notably, both compounds feature two flat bands proximate to the Fermi energy, predominantly attributed to O-$2p$ lone pairs, as evidenced by the electron localization function (ELF). Such flat bands typically exhibit weak dispersion and localized electrons, usually resulting in strong correlation effects.[20,21] The DOS analysis further unveils that the valence bands close to the Fermi level for both $Pb_{10}(PO_4)_6O$ (Figure 1b) and $Pb_{10}(PO_4)_6(OH)_2$ (Figure 1d) are predominantly influenced by O-$2p$ states, complemented by minor contributions from the Pb-($6s+6p$) states. Conversely, the conduction band is primarily derived from the Pb-($6s+6p$) states.

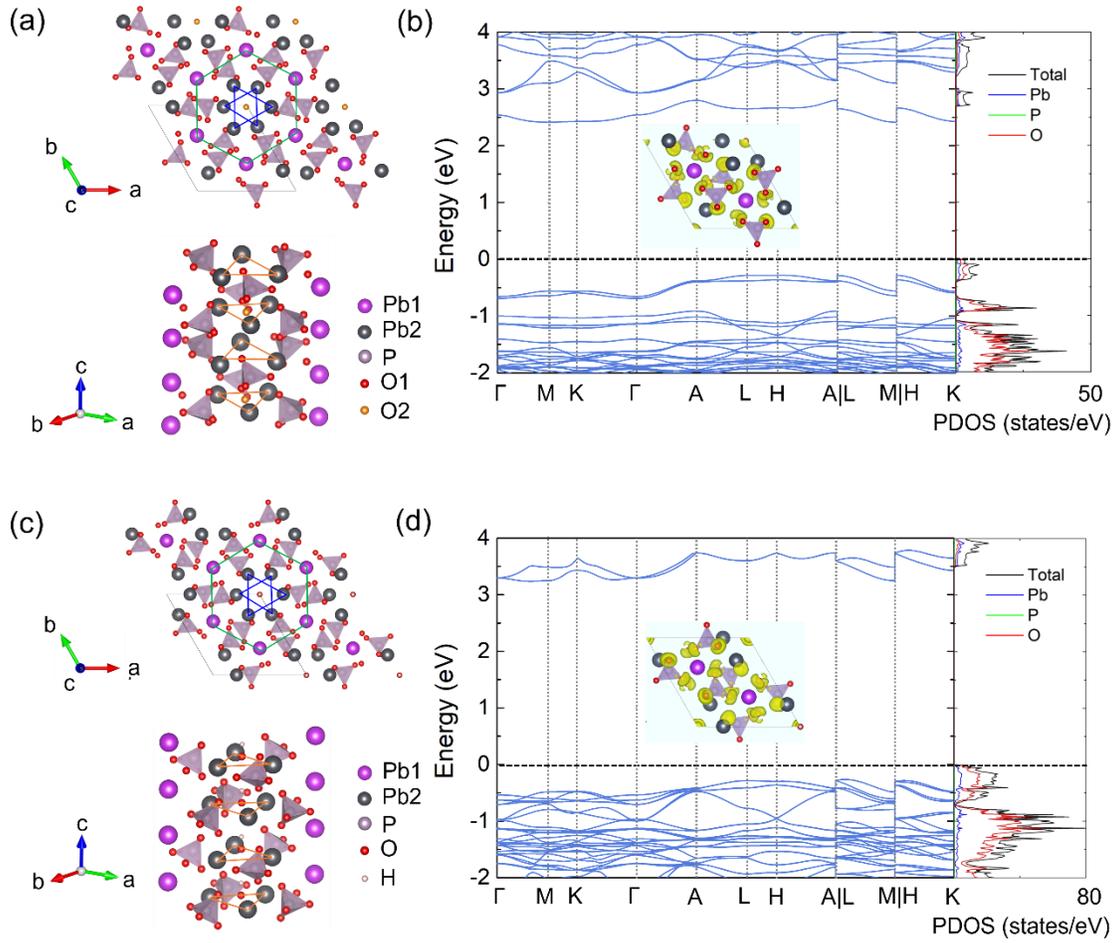

**Figure 1.** (a) The crystal structure of $Pb_{10}(PO_4)_6O$. Pb1 atoms form a hexagon marked by green lines, and Pb2 atoms form two oppositely shaped triangles marked by blue and orange lines, respectively. (b) The band structure, density of states (DOS), and electron localization function (ELF) of $Pb_{10}(PO_4)_6O$. (c) The crystal structure of $Pb_{10}(PO_4)_6(OH)_2$. (d) The band structure, DOS, and ELF of $Pb_{10}(PO_4)_6(OH)_2$.

We now turn our attention to the physical properties of Cu-doped lead apatite, i.e., LK-99.

Recent experimental findings suggest that the Pb1 site serves as the primary doping location in LK-99.[8] Concurrently, theoretical work has indicated the existence of two distinct narrow flat bands proximate to the Fermi level in the 4*f*-phase Cu-doped lead apatite. Such characteristics are postulated to be linked to superconductivity.[19,21,22] On the other hand, when doping occurs at the Pb2 site, it often introduces an isolated impurity energy band in the 6*h*-phase Cu-doped lead apatite, manifesting semiconducting-like behaviors.[20,25] Contrarily, our findings consistently highlight a semiconducting-like behavior in Cu-doped lead apatite, irrespective of the doping site, which we attribute to symmetry breaking.

Figure 2a and Figure 2c present the optimized crystal structures of 4*f*-Pb$_9$Cu(PO$_4$)$_6$O and 4*f*-Pb$_9$Cu(PO$_4$)$_6$(OH)$_2$, respectively. Unlike the trigonal *P*3 symmetry observed previously, these structures exhibit a transition to the triclinic *P*1 symmetry. This results in an energy lower of 426meV and 907meV for each compound, respectively. The derived lattice parameters for 4*f*-Pb$_9$Cu(PO$_4$)$_6$O are $a = 9.899$ Å, $b = 9.629$ Å, $c = 7.331$ Å, with angles $\alpha = 89.207°$, $\beta = 90.561°$, and $\gamma = 119.207°$, culminating in a volume of $V = 609.92$ Å$^3$. While for 4*f*-Pb$_9$Cu(PO$_4$)$_6$(OH)$_2$, the predicted lattice parameters are $a = 9.793$ Å, $b = 9.697$ Å, $c = 7.328$ Å, with angles $\alpha = 89.095°$, $\beta = 90.289°$, and $\gamma = 119.104°$, leading to a volume of $V = 607.92$ Å$^3$. Compared with non-doped Pb$_{10}$(PO$_4$)$_6$O (Pb$_{10}$(PO$_4$)$_6$(OH)$_2$), Cu doping results in a 1.0% (2.7%) volume reduction for the low-symmetry system, in contrast to 5.8% (3.9%) for the high-symmetry one. The low-symmetry system aligns more closely with experimental data indicating a 0.48% volume reduction.[8]

Noteworthy, for 4f-Pb$_9$Cu(PO$_4$)$_6$O, that while Cu-O bond lengths within the high symmetry system's CuO$_6$ octahedron are 2.07 Å and 2.28 Å within the trigonal prism, the CuO$_6$ octahedron in the low-symmetry system is notably distorted. Cu-O bond lengths range from 1.93 Å to 1.99 Å, converting into a CuO$_4$ square, as depicted in Figure 2a. Similarly, 4f-Pb$_9$Cu(PO$_4$)$_6$(OH)$_2$ showcases the CuO$_4$ square with Cu-O bond lengths from 1.95 Å to 1.98 Å, as illustrated in Figure 2c. This pronounced distortion might be linked to the semiconducting states seen in the 4f-phase Cu-doped lead apatite. Both 4f-Pb$_9$Cu(PO$_4$)$_6$O and 4f-Pb$_9$Cu(PO$_4$)$_6$(OH)$_2$ exhibit weak magnetism, registering a net magnetic momentum is 1.0 $\mu_B$ within the unit cell. The Cu ion contributes around 0.68 $\mu_B$, with residual magnetic moments credited to O atoms neighboring the Cu atom. This suggests that Cu doping substantially impacts the physical properties of lead apatite.

Figure 2b and Figure 2d display the band structure and DOS of 4f-Pb$_9$Cu(PO$_4$)$_6$O and 4f-Pb$_9$Cu(PO$_4$)$_6$(OH)$_2$, respectively. Given the inherent spin polarization in this system, we differentiate between spin up and spin down using the dashed orange line and the solid blue line within the band structure. Upon Cu doping at the Pb(1) site, an isolated impurity energy band emerges, positioned approximately 0.5 eV and 1.0 eV above the Fermi energy for 4f-Pb$_9$Cu(PO$_4$)$_6$O and 4f-Pb$_9$Cu(PO$_4$)$_6$(OH)$_2$, respectively. This is mainly primarily attributed to the Cu-3$d$ and O-2$p$ orbitals. This behavior starkly deviates from the high-symmetry (P3) 4f-phases, where two correlated half-filled flat bands align with the Fermi energy, as illustrated in Figure S1. Significantly, the band gap of 4f-Pb$_9$Cu(PO$_4$)$_6$O and 4f-Pb$_9$Cu(PO$_4$)$_6$(OH)$_2$ measures 0.84 eV and 1.31 eV, respectively, reinforcing the semiconductor nature of 4*f*-phases. The pronounced distortion within the crystal structure, particularly concerning the Cu-O frame, likely underpins the emergence of these band gaps. Consequently, our findings indicate that symmetry disruption typically inaugurates a band gap, eliminating the conspicuous flat bands at the Fermi energy level and instead presenting an isolated flat band above it.

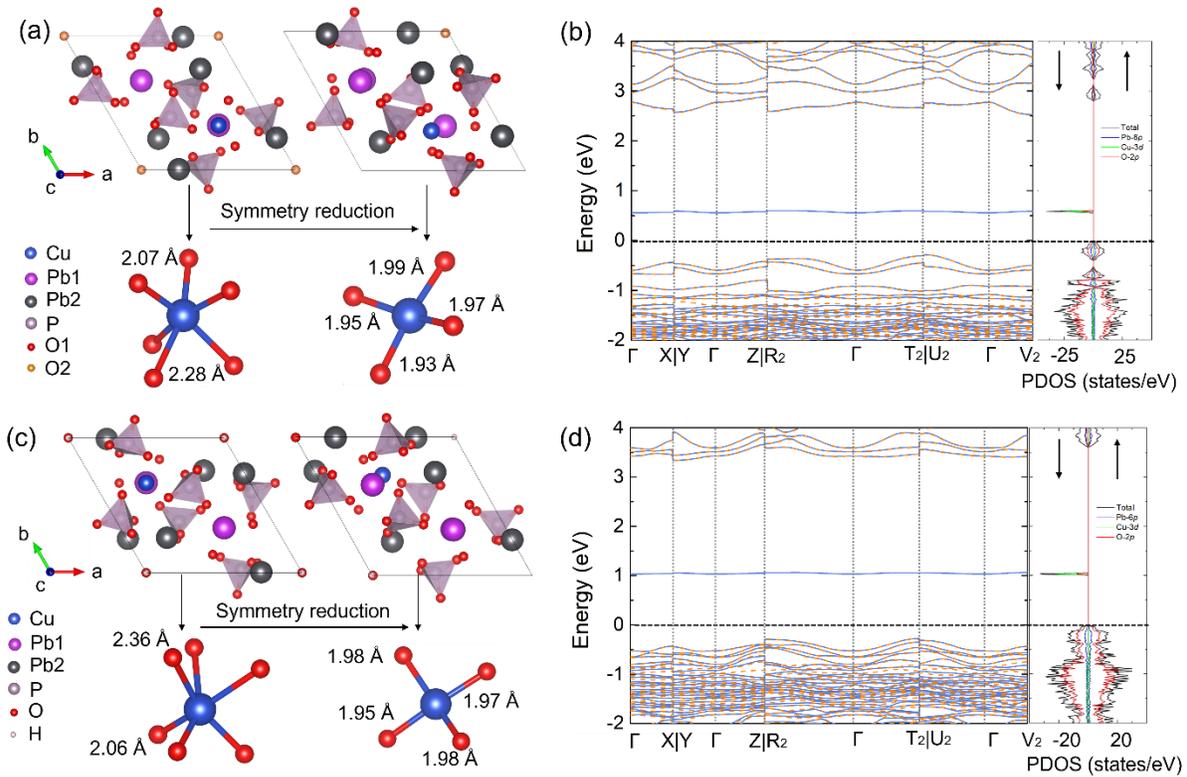

**Figure 2.** (a) The crystal structure of 4*f*-$Pb_9Cu(PO_4)_6O$. (b) The band structure and density of states (DOS) of 4*f*-$Pb_9Cu(PO_4)_6O$. (c) The crystal structure of 4*f*-$Pb_9Cu(PO_4)_6(OH)_2$. (d) The band structure and DOS of 4*f*-$Pb_9Cu(PO_4)_6(OH)_2$.

Figure 3a and Figure 3c depict the optimized crystal structure of 6$h$-Pb$_9$Cu(PO$_4$)$_6$O and 6$h$-Pb$_9$Cu(PO$_4$)$_6$(OH)$_2$, respectively. Cu doping at the Pb2 site often induces a breakdown of lattice symmetry, transitioning to the triclinic $P1$ symmetry. For 6$h$-Pb$_9$Cu(PO$_4$)$_6$O, the calculated lattice parameters are $a$ = 9.790 Å, $b$ = 9.561 Å, $c$ = 7.399 Å, $\alpha$ = 89.724°, $\beta$ = 89.896°, and $\gamma$ = 119.142°, resulting in a volume of $V$ = 604.94 Å$^3$. In the case of 6$h$-Pb$_9$Cu(PO$_4$)$_6$(OH)$_2$, the lattice parameters are $a$ = 9.724 Å, $b$ = 9.846 Å, $c$ = 7.344 Å, $\alpha$ = 90.039°, $\beta$ = 89.994°, and $\gamma$ = 120.108°, with a volume of $V$ = 608.27 Å$^3$. Relative to the non-doped lead apatite, Cu doping at the Pb2 site leads to volume reductions of 1.8% and 2.6% for the oxo and hydroxy system, respectively. The Cu-O frame in the 6$h$-phases is notably distorted, presenting a CuO$_4$ square with Cu-O bond lengths approximating 2.0 Å, as shown in Figure 3a and Figure 3c. The projected magnetic moment on the Cu atom is about 0.64 $\mu_B$, while the integrated magnetic moment for the entire unit cell is 1.0 $\mu_B$.

Significantly, the formation energy for 4$f$-Pb$_9$Cu(PO$_4$)$_6$O is lower by its 6$h$-phase counterpart by 16 meV, suggesting that the octahedrally coordinated Pb1 site is the most energetically preferred for Cu doping. Whereas Cu doping into the Pb2 site of Pb$_{10}$(PO$_4$)$_6$(OH)$_2$ emerges as more energetically favorable, exhibiting a 169meV reduction in formation energy. Given the marginal energy differences between the 4$f$-phase and the 6$h$-phase of LK-99, experimentally synthesized LK-99 samples could display inhomogeneity, possibly encompassing both 4$f$- and 6$h$-phases.

Figure 3b and Figure 3d display the band structure and DOS for the 6$h$-phase of $Pb_9Cu(PO_4)_6O$ and $Pb_9Cu(PO_4)_6(OH)_2$, respectively. The electronic structures of the $6h$-phase LK-99 demonstrate semiconducting behaviors akin to those observed in $4f$-phase LK-99. This can be attributed to the pronounced structural distortion characterized by the $P1$ symmetry. Upon shifting the doping site to Pb2, the isolated energy band repositions itself further to about 1.0 eV for $6h$-$Pb_9Cu(PO_4)_6O$ and 1.3 eV for $6h$-$Pb_9Cu(PO_4)_6(OH)_2$. This results in an enlarged band gap, measuring 1.20 eV and 1.51 eV than that of the $4f$-phase. The DOS analysis reveals the primary contributions to this band come from the Cu-$3d$ and O-$2p$ orbitals. These findings indicate that regardless of the Cu doping site, the LK-99 consistently presents the triclinic $P1$ symmetry complemented by the $CuO_4$ square coordination. This configuration facilitates the emergence of a band gap in the system, characterized by an isolated flat band, underpinning its semiconducting nature.

This study reveals that LK-99 inherently behaves like a semiconductor, a characteristic stemming from the symmetry breaking brought about by the triclinic P1 symmetry. Yet, when LK-99 is in a high-symmetry state, it often displays properties linked to superconductivity and correlated electron states. One could posit that experimentally synthesized LK-99 samples might exhibit inhomogeneity, with superconducting high-symmetry regions interspersed within non-superconducting low-symmetry zones. Consequently, the synthesis of LK-99 samples predominantly in the high-symmetry phase could pave the way to realizing room-temperature superconductors at ambient pressure.

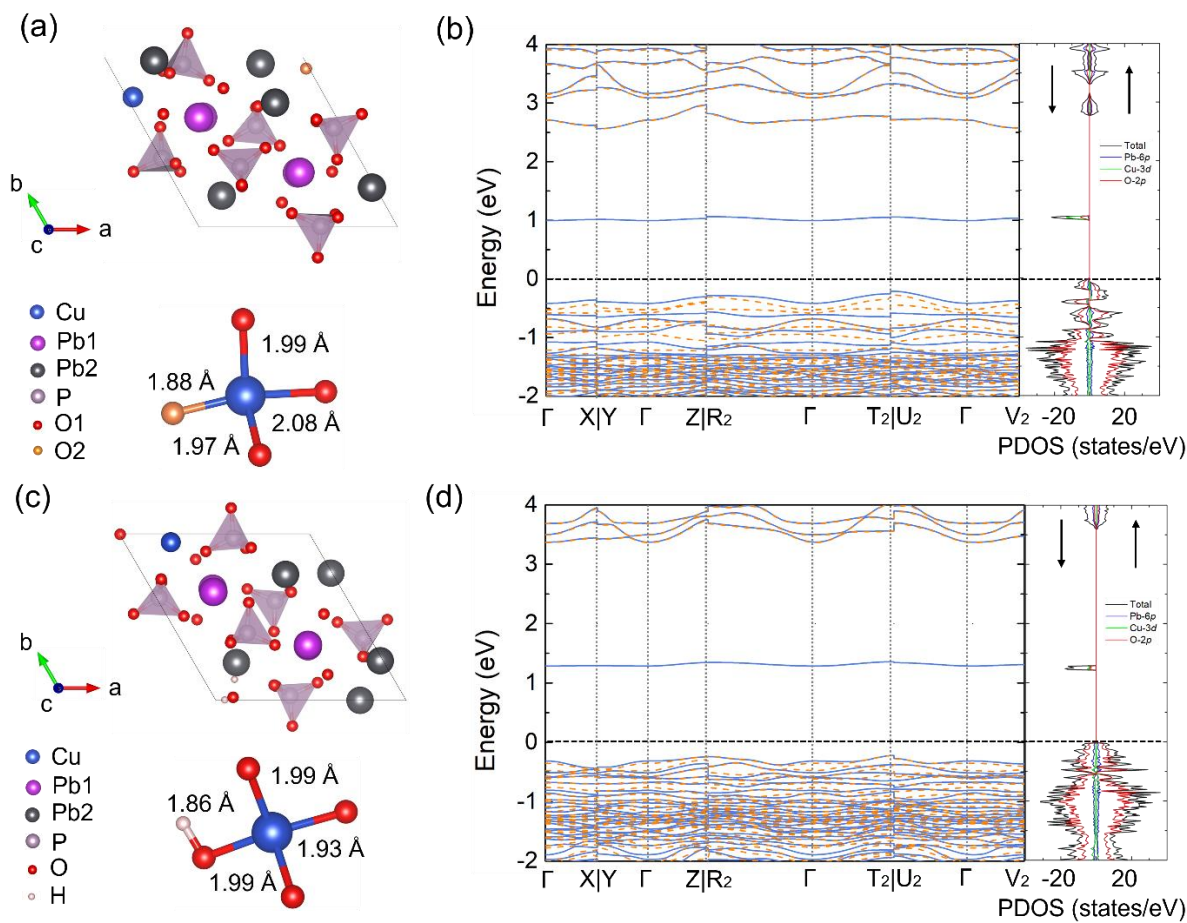

**Figure 3.** (a) The crystal structure of 6$h$-Pb$_9$Cu(PO$_4$)$_6$O. (b) The band structure and density of states (DOS) of 6$h$-Pb$_9$Cu(PO$_4$)$_6$O. (c) The crystal structure of 6$h$-Pb$_9$Cu(PO$_4$)$_6$(OH)$_2$. (d) The band structure and DOS of 6$h$-Pb$_9$Cu(PO$_4$)$_6$(OH)$_2$.

Given the detectable presence of Cu$_2$S impurity, it is plausible that S atoms are incorporated into the synthesized LK-99 samples, potentially influencing their physical properties. For the purpose of this study, we have taken 4$f$-Pb$_9$Cu(PO$_4$)$_6$O as a representative to analyze the impact of S doping. While the S doping concentration in this study is notably elevated, our primary objective is to offer deeper insight into the physical properties of LK-99 and to reconcile the discrepancies observed in experimental results related to LK-99 superconductivity.

Figure 4a and Figure 4c display the optimized crystal structure of the high-symmetry and the low-symmetry $4f$-$Pb_9Cu(PO_3S)_6O$, respectively. A mere 206 meV difference in formation energy between the two phases implies a higher chance of the low-symmetry phase present in experimental synthesis. The high-symmetry $4f$-$Pb_9Cu(PO_3S)_6O$ adopts a trigonal $P3$ symmetry, characterized by lattice parameters: $a = 10.358$ Å and $c = 7.468$ Å. This symmetry could further diminish to a triclinic $P1$ symmetry with corresponding parameters detailed in Figure 4c. The lattice parameters of $P1$ symmetry structure are $a = 10.301$ Å, $b = 10.483$ Å, $c = 7.491$ Å, $\alpha = 89.015°$, $\beta = 90.912°$, and $\gamma = 119.959°$. Compared with non-doped $Pb_{10}(PO_4)_6O$, S doping induces an approximate 13% volume increase in the system of $4f$-$Pb_9Cu(PO_3S)_6O$. Given that the theoretical volume reduction estimations for LK-99 often surpass experiment observations, the presence of S atoms in LK-99 may have been overlooked. Within the high-symmetry phase, the Cu-O bond lengths of the $CuO_6$ octahedron are 2.42 Å and 2.43 Å within the trigonal prism, while the P-S bond length extends to 2.06 Å, as demonstrated in Figure 4a. A subsequent symmetry reduction could perturb the $4f$-$Pb_9Cu(PO_3S)_6O$ phase, transitioning into a $CuO_5$ configuration with Cu-O bond lengths varying from 2.31 Å and 2.47 Å (Figure 4c). Notably, the magnetic momentum within the unit cell of the $4f$-$Pb_9Cu(PO_3S)_6O$ phase remains consistent at 1.0 $\mu_B$, primarily attributed to the Cu atom and adjacent S atoms.

Figure 4b and Figure 4d elucidate the band structures and DOS of the high-symmetry and low-symmetry variants of 4$f$-Pb$_9$Cu(PO$_3$S)$_6$O, respectively. For the high-symmetry 4$f$-Pb$_9$Cu(PO$_3$S)$_6$O, two flat bands intersect the Fermi energy level, which is similar to those of high-symmetry 4$f$-phase LK-99 (Figure 4b). It suggests the possibility of superconductivity. These two flat bands owe their presence predominately to the Cu-3$d$ and S-3$p$ orbitals. In a striking contrast, further symmetry breaking potentially generates an isolated flat band elevated by about 0.3 eV above the Fermi energy level, with the primary contribution stemming from the Cu-3$d$ and S-3$p$ orbitals (Figure 4d). It's worth noting that in comparison to 4$f$-Pb$_9$Cu(PO$_4$)$_6$O, S doping significantly narrows the band gap from 0.84 eV to 0.36 eV, advocating enhanced conductivity.

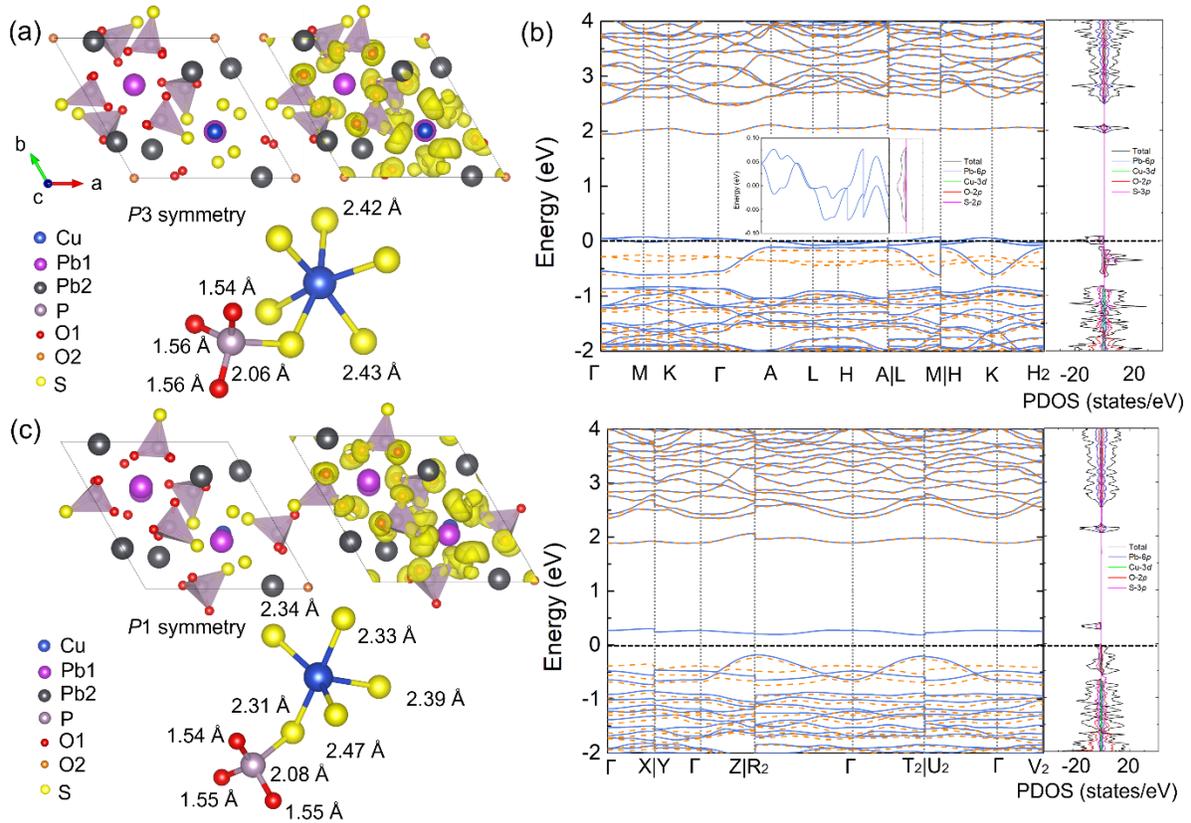

**Figure 4.** (a) The crystal structure and electronic localized function (ELF) of the high-symmetry 4f-Pb$_9$Cu(PO$_3$S)$_6$O. (b) The band structure and density of states (DOS) of 6h-Pb$_9$Cu(PO$_4$)$_6$O. (c) The crystal structure and ELF the low-symmetry 4f-Pb$_9$Cu(PO$_3$S)$_6$O. (d) The band structure, ELF, and DOS of 4f-Pb$_9$Cu(PO$_3$S)$_6$O.

In summary, we have examined the structural and electronic properties of LK-99 and its precursor compounds using QM and QM-MD simulations. Furthermore, we explore the effects of S doping on LK-99, which may arise from Cu$_2$S impurities. The findings reveal that the base compounds, Pb$_{10}$(PO$_4$)$_6$O and Pb$_{10}$(PO$_4$)$_6$(OH)$_2$, are insulators with significant band gaps. When LK-99 is doped with Cu, regardless of the site, it breaks the symmetry of the lead apatite. This leads to a notably distorted ground-state crystal structure, characterized by the triclinic $P1$ symmetry, accentuated by the CuO$_4$ square coordination. Consequently, LK-99 often demonstrates semiconducting properties, characterized by an isolated flat band above the Fermi energy level. This band primarily comes from the Cu-3$d$ and O-2$p$ orbitals. Similarly, when S is introduced to LK-99, it maintains the triclinic $P1$ symmetry, but the resulting flat band has a significantly reduced band gap. This band prominently originates from the Cu-3$d$ and S-3$p$ orbitals. These insights not only shed light on the inherent properties of LK-99 but also offer valuable direction for crafting high-$T_C$ superconductors.

## ASSOCIATED CONTENT

**Supporting Information**

(i) The computational details of quantum mechanics (QM) and quantum mechanics molecular dynamics (QM-MD) simulations; (ii) Figure S1a displays the band structure and density of states (DOS) of the high-symmetry ($P3$) 4$f$-Pb$_9$Cu(PO$_4$)$_6$O, and Figure S1b displays the band structure and DOS of the high-symmetry ($P3$) 4$f$-Pb$_9$Cu(PO$_4$)$_6$(OH)$_2$;

## AUTHOR INFORMATION

**Corresponding Author**

**Qi An** -- *Department of Materials Science and Engineering, Iowa State University, Ames,*

*Iowa 50011, United States*; orcid.org/ 0000-0003-4838-6232; Email: qan@iastate.edu.

**Author**

**Jun Li** -- *Department of Materials Science and Engineering, Iowa State University, Ames, Iowa 50011, United States*; orcid.org/0000-0002-0209-8043.

**Author Contributions**

QA conceived and designed the simulations. JL constructed the simulation model and performed the simulations. All authors provided insights into data analysis and wrote the manuscript. All authors have given approval to the final version of the manuscript.

**Notes**

The authors declare no competing financial interest. While preparing this manuscript, we became aware of DFT studies using Quantum Espresso on arXiv, which also shows symmetry breaking in the superconductivity of LK-99.[28,29] Our research was conducted independently and did not reference these studies.

**ACKNOWLEDGMENTS**

This work was supported by the start-up research grant at Iowa State University.